\documentclass{ECOC-author}

\usepackage{tikz}
\usetikzlibrary{shapes,arrows,positioning,fit}
\usepackage{xcolor}
\definecolor{rasOrange}{rgb}{1,0.7451,0.0039}
\begin{document}
\title{End-to-end Learning for GMI Optimized Geometric Constellation Shape}

\author{Rasmus T Jones\ad{1}, Metodi P Yankov\ad{1,2}\corr\, and Darko Zibar\ad{1} }

\address{\add{1}{Technical University of Denmark, DTU Fotonik, Kgs. Lyngby, Denmark}\add{2}{Fingerprint Cards A/S, 2730 Herlev, Denmark}
\email{meya@fotonik.dtu.dk}}


\begin{abstract}
Autoencoder-based geometric shaping is proposed that includes optimizing bit mappings. Up to 0.2 bits/QAM symbol gain in GMI is achieved for a variety of data rates and in the presence of transceiver impairments. The gains can be harvested with standard binary FEC at no cost w.r.t. conventional BICM. 
\end{abstract}

\maketitle

\section{Introduction}
Constellation shaping is now a well-established technique in the optical communications research community for improving the system throughput and operate close to the theoretically achievable information rate and at high spectral efficiency.
Both probabilistic shaping \cite{b01, b02, b03} and geometric shaping \cite{b06, b07, b08, b10, b12}, each having its benefits and drawbacks, are of significant interest to both system vendors and researchers. 
Probabilistic shaping refers to optimizing the probabilities of the constellation points, typically applied to rectangular quadrature amplitude modulation (QAM), whereas geometric shaping refers to optimizing the positions of the constellation points.

Probabilistic shaping provides slightly higher shaping gains than geometric shaping in terms of the generalized mutual information (GMI), since non-rectangular constellations lack an obvious Gray-like code. This problem can be avoided, in trade-off for increased complexity, by employing Geometric shaping with iterative demapping and decoding~\cite{b07}, or with non-binary forward error correction (FEC)~\cite{b06}.
Also probabilistic shaping includes similar trade-off, as it requires a shaping encoder, such as the constant composition distribution matcher~\cite{b03,b04,b05}. Such encoders may be problematic due to their error-propagation and inherent serialized processing, making it challenging for parallelized implementation.
Hence, a geometric constellation shape with Gray-like code combines low implementation complexity and shaping gain. Such constellations have been proposed in~\cite{b10}, yet the applied method switches in between two independent optimization methods and is not scalable in a straight forward manner.

In this paper, an autoencoder is used to optimize a geometric shaped constellation in terms of GMI by jointly optimizing the bit mapping and the position of the constellation points.
The proposed method makes no assumption on the channel model
and is easily scalable to constellations of higher order and higher dimension. The optimization builds upon the autoencoder-based geometric shaping, optimized for mutual information (MI) only \cite{b12}. The following myths about geometric shaping are debunked with the proposed method:
\begin{enumerate}
  \item \textit{Conventional bit-interleaved coded modulation (BICM) is penalized with geometric shaping due to the non-Gray labeling.} We show that the proposed autoencoder arrives at a Gray-like code, which does not exhibit this problem.
  \item \textit{The implementation penalty is higher for geometric shaping than rectangular QAM.} We show that in the operating regions of interest and with the application of modulation-format independent digital signal processing (DSP) chain, the penalty is the same.
  \item \textit{Iterative demapping or non-binary FEC are required for geometric shaping schemes.} We show that the proposed labelings do not have this requirement because they are GMI optimized.
\end{enumerate}

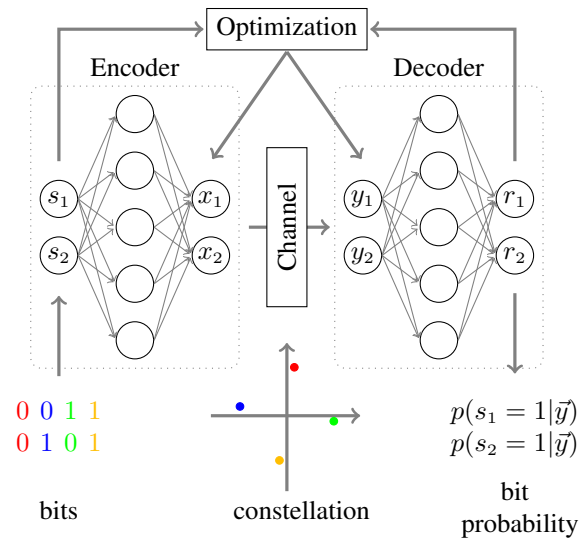
\begin{figure}[t]
	\centering\begin{tikzpicture}[shorten >=1pt,->,draw=black!50, node distance=\layersep]
	\tikzstyle{block} = [rectangle,draw=black, text centered, minimum width=6em, minimum height=1.5em]
    \tikzstyle{neuron}=[circle,fill=white,draw=black,minimum size=14pt,inner sep=0pt]
    \tikzstyle{bias}=[circle,dashed,fill=gray!20,draw=black,minimum size=12pt,inner sep=0pt]
    \tikzstyle{annot} = [text width=4em, text centered]
	\tikzstyle{line} = [draw, --]
	\tikzstyle{arrow} = [draw, -latex']
	
	% arrays
	\def\names{{"A","I","Q"}}%
	\def\xnames{{"A","$x_1$","$x_2$"}}%
	\def\ynames{{"A","$y_1$","$y_2$"}}%
	\def\inputnames{{"A","$s_1$","$s_2$","$s_3$","$s_4$"}}% 
	\def\outputnames{{"A","$r_1$","$r_2$","$r_3$","$r_4$"}}% 
	\def\outputnamesRight{{"A","$p(s_1=1|{}\cdot{})$","$p(s_2=1|{}\cdot{})$","$p(s_3=1|{}\cdot{})$","$p(s_4=1|{}\cdot{})$"}}% 
    
	% Draw the input layer nodes
    \foreach \name / \y in {1,...,2}
        \path[yshift=-0.75cm]
        node[neuron] (eI-\name) at (0,-0.75*\y) {\pgfmathparse{\inputnames[\y]}\pgfmathresult};
    % Draw the hidden layer nodes
    \foreach \name / \y in {1,...,5}
        \path[yshift=0.375cm]
            node[neuron] (eH-\name) at (1.00cm,-0.75*\y) {};	
    % Draw the output layer node
    \foreach \name / \y in {1,...,2}
    	\path[yshift=-0.75cm]
    		node[neuron] (eO-\name) at (2.00cm,-0.75*\y) {\pgfmathparse{\xnames[\y]}\pgfmathresult};
    		
	% Connect every node in the input layer with every node in the
    % hidden layer.
    \foreach \source in {1,...,2}
        \foreach \dest in {1,...,5}
            \path (eI-\source.east) edge (eH-\dest.west);
    % Connect every node in the hidden layer with the output layer
    \foreach \source in {1,...,5}
    	\foreach \dest in {1,...,2}
        	\path (eH-\source.east) edge (eO-\dest.west);
	
	\coordinate[left of=eH-3, node distance=1.75cm] (arrow1Start);	
	\coordinate[left of=eH-3, node distance=1.25cm] (arrow1End);
	%\draw[very thick] (arrow1Start) -- (arrow1End);
	
	%\coordinate[right of=eH-3, node distance=1.25cm] (arrow2Start);	
	%\coordinate[right of=eH-3, node distance=1.75cm] (arrow2End);
	%\draw[very thick] (arrow2Start) -- (arrow2End);
	
	%\node[right of=eH-3,node distance=3.125cm] (points)
	%	{\includegraphics[width=0.125\linewidth]{points.pdf}};
		
	\coordinate[right of=eH-3, node distance=1.5cm] (arrow3Start);	
	\coordinate[right of=eH-3, node distance=2.6cm] (arrow3End);
	\draw[very thick] (arrow3Start) -- (arrow3End);
	
	\node [block,right of=eH-3,node distance=2cm,fill=white,rotate=90] (channel) {Channel};
	
	\coordinate[below of=channel, node distance=3.5cm] (yAxisStart);
    \coordinate[below of=channel, node distance=1.5cm] (yAxisStop);
    \draw[very thick] (yAxisStart) -- (yAxisStop);
	
	\coordinate[below of=channel, node distance=2.5cm] (yAxisMid);
	\coordinate[left of=yAxisMid, node distance=1.0cm] (xAxisLeft);
	\coordinate[right of=yAxisMid, node distance=1.0cm] (xAxisRight);
	\draw[very thick] (xAxisLeft) -- (xAxisRight);
	
	\coordinate[left of=yAxisMid, node distance=0.5cm] (xAxisLeftMid);
	
	\draw[red,fill=red] (3.0978,-3.7327) circle (0.5mm);
	\draw[blue,fill=blue] (2.3827,-4.2522) circle (0.5mm);
	\draw[green,fill=green] (3.6173,-4.4478) circle (0.5mm);
	\draw[rasOrange,fill=rasOrange] (2.9022,-4.9673) circle (0.5mm);
	%\node[right of=channel,node distance=2cm] (pointsDecision)
	%	{\includegraphics[width=0.125\linewidth]{pointsDecisionTest.pdf}};
		
	%\coordinate[right of=channel, node distance=1.375cm] (arrow4Start);	
	%\coordinate[right of=channel, node distance=1.875cm] (arrow4End);
	%\draw[very thick] (arrow4Start) -- (arrow4End);
		
	% Draw the input layer nodes
    \foreach \name / \y in {1,...,2}
    	\path[yshift=-0.75cm]
        	node[neuron] (dI-\name) at (4cm,-0.75*\y) {\pgfmathparse{\ynames[\y]}\pgfmathresult};
	% Draw the hidden layer nodes
    \foreach \name / \y in {1,...,5}
        \path[yshift=0.375cm]
            node[neuron] (dH-\name) at (5cm,-0.75*\y) {};	
	% Draw the output layer node
    \foreach \name / \y in {1,...,2}
	\path[yshift=-0.75cm]
	node[neuron] (dO-\name) at (6cm,-0.75*\y) {\pgfmathparse{\outputnames[\y]}\pgfmathresult};

	% \node[bias] (bias-1) at (0.35cm,-3.5cm) {\footnotesize 1};
	% \node[bias] (bias-2) at (1.5cm,-3.5cm) {\footnotesize 1};

 %    \foreach \dest in {1,...,5}
 %            \path (bias-1.east) edge (eH-\dest.west);
 %    \foreach \dest in {1,...,2}
 %            \path (bias-2.east) edge (eO-\dest.west);
    
	% Connect every node in the input layer with every node in the
    % hidden layer.
    \foreach \source in {1,...,2}
        \foreach \dest in {1,...,5}
            \path (dI-\source.east) edge (dH-\dest.west);
    % Connect every node in the hidden layer with the output layer
    \foreach \source in {1,...,5}
    	\foreach \dest in {1,...,2}
        	\path (dH-\source.east) edge (dO-\dest.west);
        	
    \coordinate[below of=eI-2, node distance=0.5cm] (bitsLeftStart);
    \coordinate[below of=eI-2, node distance=1.6cm] (bitsLeft);
    \draw[very thick] (bitsLeft) -- (bitsLeftStart);
    
    \coordinate[below of=dO-2, node distance=0.5cm] (bitsRightStart);
    \coordinate[below of=dO-2, node distance=1.6cm] (bitsRight);
    \draw[very thick] (bitsRightStart) -- (bitsRight);
    
    \node[node distance=0.7cm, below of=bitsRight,yshift=0.0cm] (bitsOut) {
    $   \begingroup % keep the change local
			\setlength\arraycolsep{2pt}
			\begin{matrix}
    p(s_1=1|\vec{y}) \\
    p(s_2=1|\vec{y})
			\end{matrix}
		\endgroup
	$
	};
    
    \node[annot,below of=channel, node distance=3.75cm] (constLabel) {constellation};
    \node[annot,left of=constLabel, node distance=3cm] (bitLabel) {bits};
    \node[annot,right of=constLabel, node distance=3cm] (probLabel) {bit\\ probability};
    
    \node[node distance=0.7cm, below of=bitsLeft,yshift=0.0cm] (bitsIn) {
    $   \begingroup % keep the change local
			\setlength\arraycolsep{2pt}
			\begin{matrix}
    {\color{red}0} & {\color{blue}0} & {\color{green}1} & {\color{rasOrange}1} \\
    {\color{red}0} & {\color{blue}1} & {\color{green}0} & {\color{rasOrange}1}
			\end{matrix}
		\endgroup
	$
	};
	
	\node [block,above of=channel, node distance=2.625cm] (opt) {Optimization};
	\coordinate[above of=eI-1, node distance=0.5cm] (optLeftStart);
	\coordinate[above of=eI-1, node distance=2.25cm] (optLeft);
	\draw[very thick] (optLeftStart) -- (optLeft) -- (opt);

	\coordinate[above of=dO-1, node distance=0.5cm] (optRightStart);
	\coordinate[above of=dO-1, node distance=2.25cm] (optRight);
	\draw[very thick] (optRightStart) -- (optRight) -- (opt);
	
	\coordinate[below of=opt, node distance=0.3cm] (optTop);
	\coordinate (optTopLeft) at (2.0cm,-1cm);
	\draw[very thick] (optTop) -- (optTopLeft);
	
	\coordinate (optTopRight) at (4.0cm,-1cm);
	\draw[very thick] (optTop) -- (optTopRight);
	
	\node[draw,dotted,rounded corners,fit=(eI-1) (eI-2) (eH-1) (eH-2) (eH-3) (eH-4) (eH-5) (eO-1) (eO-2)] (labelOne) {};
	\node[annot,above of=eH-1, node distance=0.625cm] (il) {Encoder};
	
	\node[draw,dotted,rounded corners,fit=(dO-1) (dO-2) (dH-1) (dH-2) (dH-3) (dH-4) (dH-5) (dI-1) (dI-2)] (labelTwo) {};
	\node[annot,above of=dH-1, node distance=0.625cm] (ol) {Decoder};
	
\end{tikzpicture}
    \caption{Schematic autoencoder model applied directly on bits.}
    \label{fig0}
\end{figure}

\section{Methology}
\subsection{Autoencoder}
Similar to the setup in~\cite{b12}, an autoencoder with embedded channel model is trained for a geometric constellation shape of order $M$:
\begin{equation}
\begin{split}
    \vec{x} &= f(\vec{s}),\\
    \vec{y} &= c(\vec{x},P^3,\kappa,\kappa_3),\\
    \vec{r} &= g(\vec{y})
    \end{split}
\end{equation}
where $f(\cdot)$ is the encoder neural network, $c(\cdot)$ is a fiber channel model and $g(\cdot)$ is the decoder neural network. \begin{figure}[t]
        \centering\includegraphics[width=0.60\linewidth]{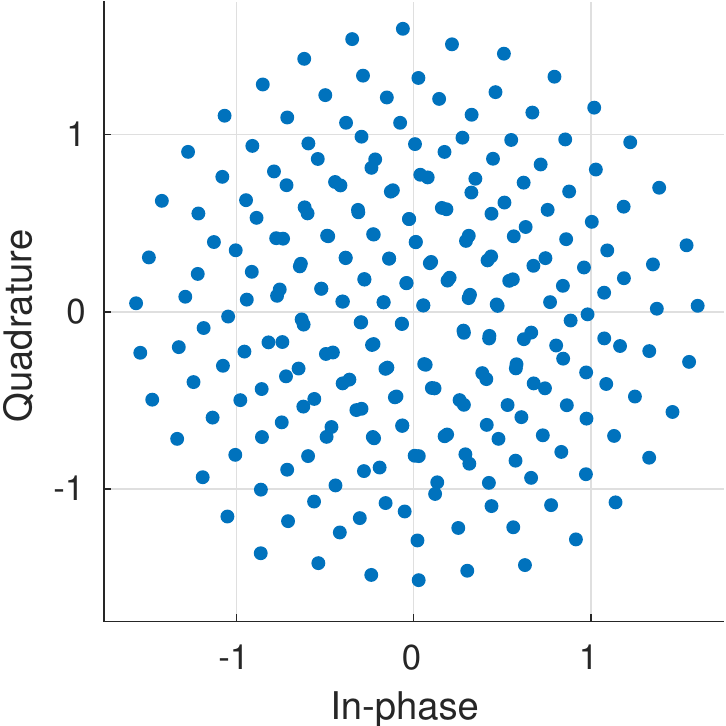}
        \centering\includegraphics[width=0.60\linewidth]{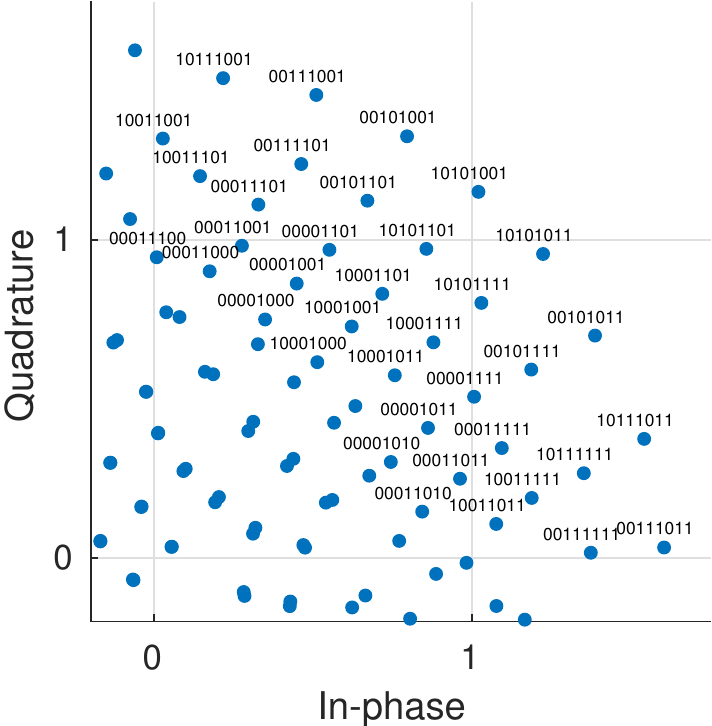}
\caption{Geometric constellation shape of order $M$=256~\textbf{(top)} and a zoomed-in version~\textbf{(bottom)} including bit mappings.}
\label{fig1}
\end{figure}
The used fiber channel model~\cite{b11} is dependent on the launch power $P$, and on the fourth and sixth order moment of the transmitted constellation, $\kappa$ and $\kappa_3$. The channel input $\vec{x}$ represents samples from a constellation and the channel output $\vec{y}$ the respective observations.

In contrast to the setup in~\cite{b12}, where the autoencoder is trained with a dataset of one-hot encoded vectors and softmax output layer, here the input space is $\vec{s} \in \mathcal{S}=\{0,1\}^m$, the set of all possible bit sequences of length $m=\log_2{M}$. Accordingly, the target space is given by
$\vec{r} \in \mathcal{R} = \{x~\in~\mathbb{R} | 0<x<1\}^m$, which refers to an output layer with sigmoid activation function, and represents the posterior probabilities of the bits being 1 or 0, $p(\vec{s}|\vec{y})$.
A training batch of size~$K$,
$\{ \vec{s}^{(1)}, \vec{s}^{(2)}, ..., \vec{s}^{(K)}\}$,
is uniformly sampled from $\mathcal{S}$ and propagated through the autoencoder.
The weights of the autoencoder are updated with stochastic gradient descent according to a loss function:
\begin{equation}
\begin{split}
    & L(\{ \vec{s}^{(1)}, ..., \vec{s}^{(K)}\}) = \frac{1}{K} \sum_{k=1}^{K} l(\vec{s}^{(k)},\vec{r}^{(k)}),\\
    & l(\vec{s},\vec{r}) =-\frac{1}{m} \sum_{i=1}^{m} s_i \log(r_i) + (1 - s_i) \log(1 - r_i).
\end{split}
\end{equation}
Since the input and output vectors represent bit sequences, the encoder neural network is learning both a geometric constellation shape and a bit mapping rule.
Following the mismatched decoding principle (described in e.g. \cite{b01}), the log-likelihood loss in Eq. (2) is an upper bound to the conditional entropy $L \cdot m \ge {\mathcal{H}}(\mathbf{S}|\mathbf{Y})$ and minimizing it thus leads to the maximization of the $\text{GMI}={\mathcal{H}}(\mathbf{S})-{\mathcal{H}}(\mathbf{S}|\mathbf{Y})$, where ${\mathcal{H}}(\mathbf{S})=m$ is the entropy of the input.
%The log-likelihood loss in Eq. (2) is directly proportional to the conditional entropy ${\mathcal{H}}(\mathbf{S}|\mathbf{R})$, and minimizing it thus leads to the maximization of the GMI, given by $\text{GMI}={\mathcal{H}}(\mathbf{S})-{\mathcal{H}}(\mathbf{S}|\mathbf{R})$, where ${\mathcal{H}}(\mathbf{S})$ is the entropy of the input, and $\mathbf{R}$ is the output.  
In Fig.~\ref{fig1}~(top), an example constellation of order $M$=256 is shown, with a zoomed-in version~(bottom), illustrating that a Gray-like bit mapping is achieved (only a selected subset of labels shown for clarity). An implementation of the presented autoencoder is available online as Python/TensorFlow programs~\cite{b13}.
\begin{figure}[b]
\centering\includegraphics[width=0.93\linewidth]{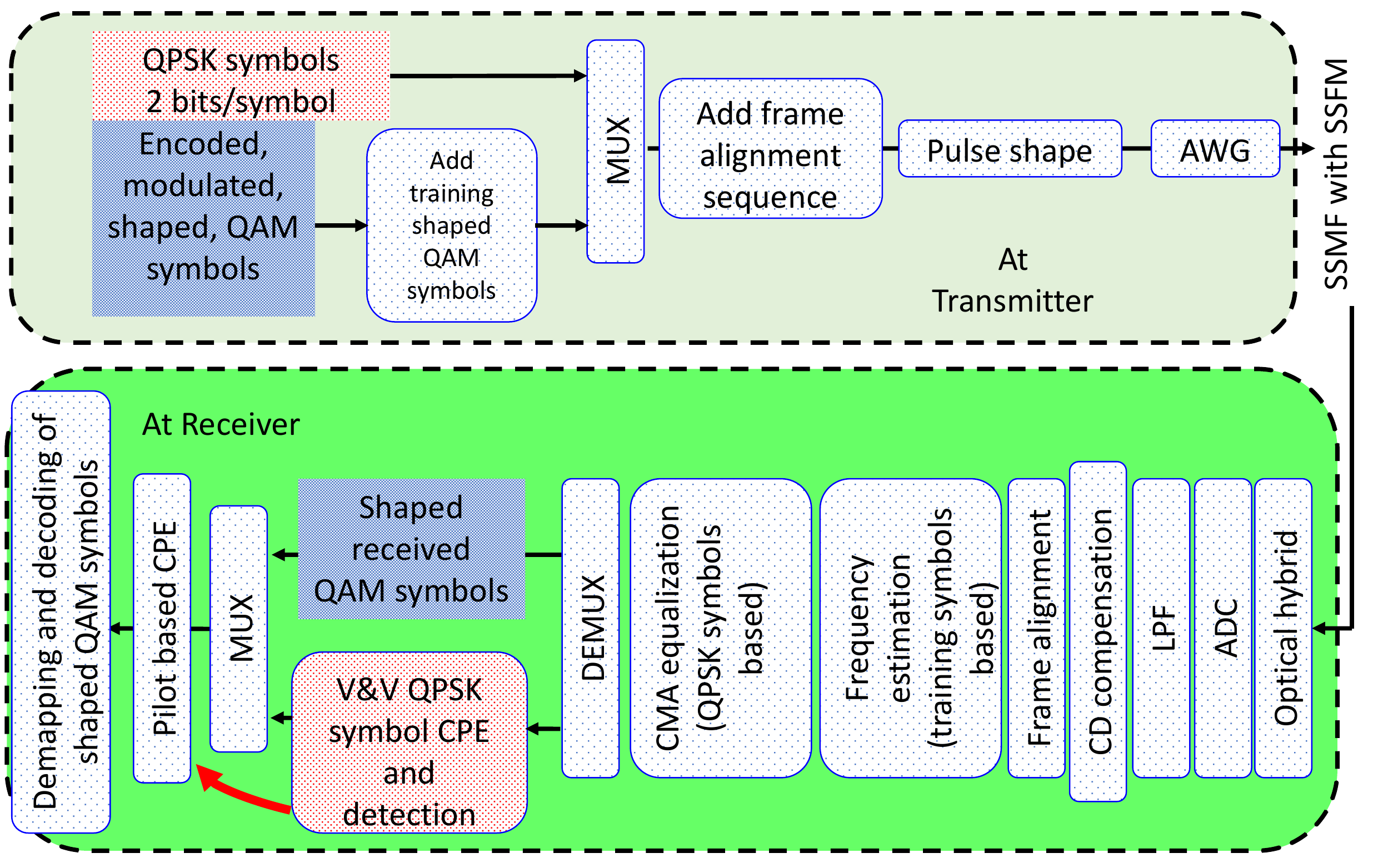}
\caption{Setup of pilot based system including transceiver impairments.}
\label{fig2}
\end{figure}

\subsection{System simulation and DSP chain}
A QPSK-hybrid quasi-pilot-based DSP chain is considered which also takes advantage of the distribution of the modulation format. The system is adopted from~\cite{b09}, and is briefly summarized below and in Fig.~\ref{fig2} for completeness. The FEC-encoded modulation symbols are interleaved with 10\% QPSK symbols, which carry data at rate 2 bits/symbol. A Zaddoff-Chu pre-amble sequence is inserted before the 1st transmitted block for frame synchronization. Square-root raised cosine pulse shape is applied at the transmitter with a roll-off factor of 0.01. The entire waveform is sent on a standard single mode fiber ($\alpha = 0.2 \frac{dB}{km}, D = 17 \frac{ps}{nm \cdot km}, \gamma = 1.3 \frac{1}{W \cdot km}$), simulated with the split-step Fourier method. At the receiver, the pre-amble is used to detect the start of transmission. Then the QPSK symbols are detected with the Viterbi\&Viterbi method. The high spectral efficiency operating point easily results in error-free QPSK symbols, which after detection are used as pilots for two sample per symbol CMA equalization and one sample per symbol carrier phase estimation (CPE). After phase tracking, MI is used to estimate the achievable information rate, the symbols are demapped and the soft bits are sent for FEC decoding. The autoencoder decoder part $g(\vec{y})$ is only used for the optimization and is replaced by classical Gaussian auxiliary channel for the actual MI and GMI estimation and FEC decoding as in~\cite{b01, b09}. Iterative demapping and decoding is not performed. The FEC is the LTE standard turbo code \cite{b14}. The data rate is controlled easily by puncturing the FEC. For example, a data rate of 6~bits/QAM symbol is achieved with 256QAM by puncturing the rate R=$\frac{1}{3}$ mother code to a new code rate of R=$\frac{3}{4}$, corresponding to 33\% overhead. For simplicity, a data rate step of 0.5~bits/symbol is considered in this paper, although a smaller step is straight forward to achieve.

\section{Results}
The proposed designs are evaluated with a wavelength division multiplexed (WDM) system of 5 channels, 20~GBd each, with 22~GHz spacing. An Erbium doped fiber amplifier-based link is considered with amplifier spacing of 80~km and noise figure of 5~dB. At each distance, the launch power is swept and the performance at the optimal launch power is reported in each case for the central WDM channel. For simpler comparisons, the data rate achieved by the QAM symbols is reported only, and the constant (for all formats) addition of the 2~bit/symbol 10\% QPSK symbols is not shown.
\begin{figure*}[t]%[htdp]
    \centering
    \begin{minipage}{.5\textwidth}
        \centering
        \includegraphics[width=0.92\linewidth]{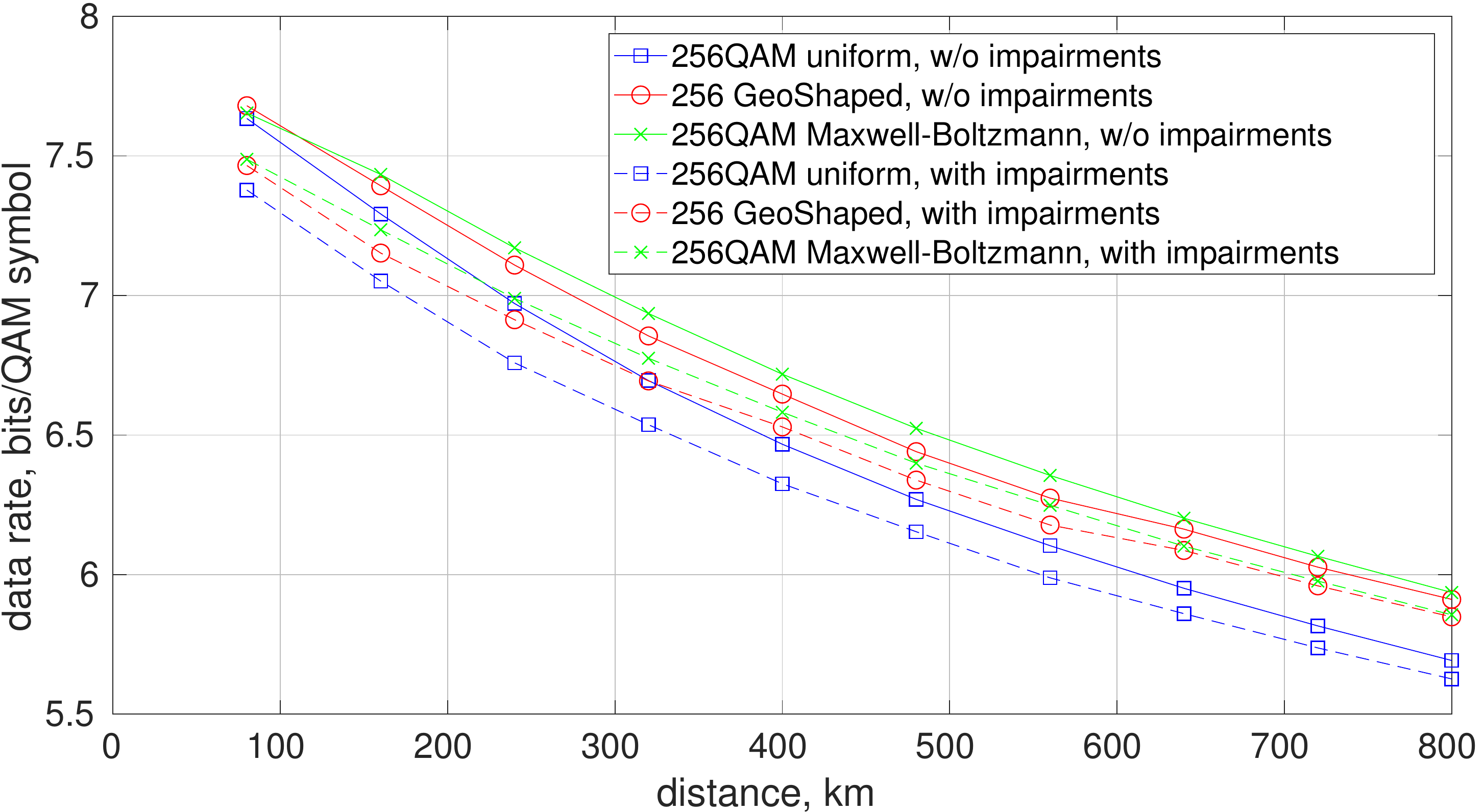}
    \end{minipage}%
    \begin{minipage}{0.5\textwidth}
        \centering
        \includegraphics[width=0.92\linewidth]{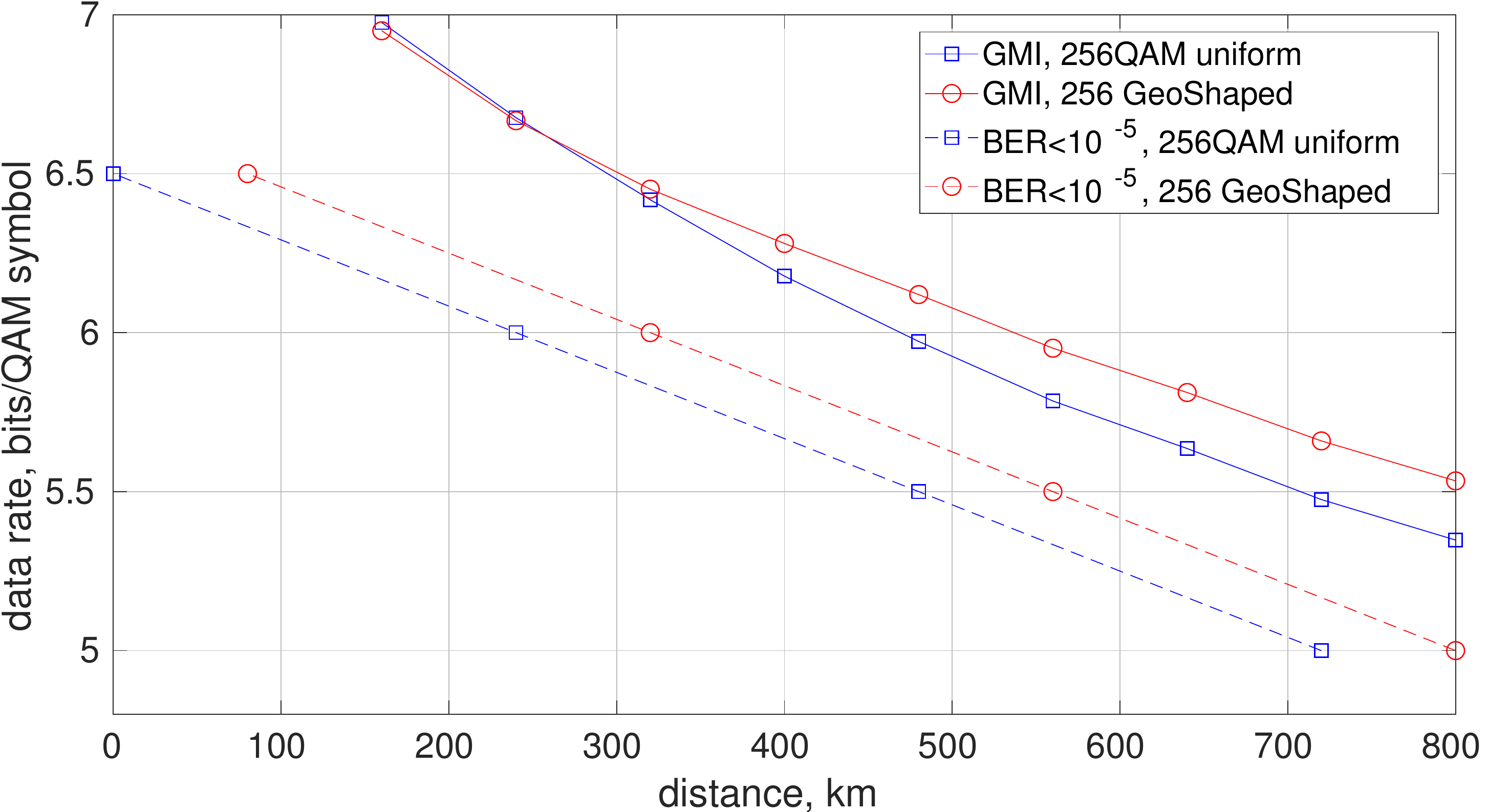}
    \end{minipage}
    \caption{\textbf{(left)} MI vs. distance for the studied modulation formats. Geometric shaping achieves similar gain to probabilistic shaping, both with and without impairments. \textbf{(right)} GMI and achieved error-free data rates as a function of the transmission distance with the impairments on. The shaping gain is 1 span, and comes for free.}
    \label{fig3}
\end{figure*}
In order to evaluate the geometric shaping method in a realistic environment, the following impairments are introduced at the receiver and transmitter: 1) laser linewidth of 10~kHz, modelled with a Wiener process; 2) frequency offset between transmitter laser and local oscillator of 50~MHz; 3) ADC sampling frequency of 80~GSa/s; 4) ADC resolution of 6~bits, modelled with a uniform quantization step. A family of three geometric shapes optimized as in Section 2 are evaluated. The shapes are optimized for transmission at 2, 5 and 10 spans. The optimal of the three is reported at each distance. For reference, a standard, Gray-coded 256QAM and a probabilistic shaped Gray-coded 256QAM achieved with the Maxwell-Boltzmann probability mass function \cite{b02} are studied. The latter is optimized at each distance and each launch power. The MI at the optimal launch power of all formats is given in Fig.~\ref{fig3}~(left) as a function of the transmission distance both with (solid lines) and without (dashed lines) impairments. 
Geometric shaping achieves between $\approx$~50\% and  $\approx$~90\% of the probabilistic shaping gain, both with and without impairments, demonstrating that geometric shapes are not penalized more than rectangular  constellations by the transceiver impairments, provided that a proper DSP is employed. Up to 0.2 bits/symbol of implementation penalty can be noticed for all formats, slightly more pronounced at the short distances/high rates. It should be noted that an efficient transmission system should switch to a larger constellation for those operating points, for which the quantization loss is no longer the dominant effect, but rather the linear and nonlinear transmission noise.
Finally, the effects of the mapping on the total achievable throughput is studied in Fig.~\ref{fig3}~(right) with the impairments on. The GMI at the optimal launch power is reported with solid lines, and \textit{the maximum distance, at which error-free transmission was achieved at the optimal launch power for a given input data rate} is given in dashed lines. As seen, the GMI gain is slightly lower (0.2~bits/symbol) than the MI gain (up to 0.3~bits/symbol) and is translated to one span of achieved error-free distance. Since there is no change to the BICM structure, the demapper and decoder, but only to the mapping function, this gain comes for free.

\section{Conclusion}
Autoencoder based GMI optimization of geometric shapes and mapping functions was proposed for high spectral efficiency WDM systems. Neither changes to the BICM coding structure nor iterative demapping and/or non-binary FEC are required in order to achieve shaping gain. The shaping gain of 0.2~bits/symbol or alternatively 1~span of transmission therefore comes for free, including in the presence of typical transceiver impairments, such as ADC quantization and limited sampling frequency, as well as laser phase noise.

\begin{footnotesize}
%\section*{Acknowledgements
\textbf{Acknowledgements}
This work was financially supported by the European Research Council through the ERC-CoG FRECOM project (grant agreementno. 771878).
\end{footnotesize}

\end{document}